\documentclass[10pt,aps,prl,twocolumn,groupedaddress]{revtex4-1}%
\usepackage{graphicx}
\usepackage[per-mode=fraction, range-phrase = -, range-units = single, exponent-product =\cdot, inter-unit-product = \cdot, separate-uncertainty = true]{siunitx}
\usepackage[hidelinks]{hyperref}
\usepackage{regexpatch}%
\usepackage{amsmath}%
\usepackage{amsfonts}%
\usepackage{amssymb}

\begin{document}

\title{Fast in situ observation of atomic Feshbach resonances by photoassociative ionization}
\author{M. Eisele}\email[]{max-albert.eisele@uni-tuebingen.de}
\author{R.A.W. Maier} 
\author{C. Zimmermann}
\affiliation{Physikalisches Institut, Eberhard Karls Universit\"{a}t T\"{u}bingen, Auf der Morgenstelle 14, D-72076 T\"{u}bingen, Germany}

\date{\today}

\begin{abstract}
We propose and experimentally investigate a scheme for observing Feshbach resonances in atomic quantum gases in situ and with a high temporal resolution of several ten nanoseconds. 
The method is based on the detection of molecular ions, which are optically generated from atom pairs at small interatomic distances. 
As test system we use a standard rubidium gas ($^{87}$Rb) with well known magnetically tunable Feshbach resonances.
The fast speed and the high sensitivity of our detection scheme allows to observe a complete Feshbach resonance within one millisecond and without destroying the gas.
\end{abstract}
\maketitle

Many exciting phenomena in atomic quantum gases are closely associated with a change in the atomic correlation functions.
Some prominent examples are the Mott insulator transition \cite{Greiner2002}, the BEC-BCS crossover \cite{Regal2004,*Zwierlein2004,*Bartenstein2004,*Bourdel2004}, few body physics based on magnetic Feshbach resonances \cite{Chin2010}, or more recently, experiments on unitary quenches and Tan's contact intensities \cite{Fletcher2017}. 

Despite its obvious importance, up to now, second order correlations have not been observed directly with high temporal resolution and without destroying the gas.
Usually they are observed indirectly for instance by monitoring three-body losses:
If the probability of finding two atoms at close distance is enhanced, the collision with a third atom may induce the formation of a molecular dimer plus an atom that carries away the binding energy.
After each experimental preparation cycle, the remaining atom number is monitored for various interaction times.
From the resulting loss curve the three-body loss coefficient is extracted.
An experimental cycle includes cooling, trapping, manipulating and detection of the atoms and typically lasts several ten seconds. 
The determination of a single loss coefficient thus may take an hour or longer depending on the desired statistical quality. 
Feshbach type scattering resonances \cite{Chin2010} and Efimov three-body states \cite{Naidon2017} have been observed this way in heroic efforts with continuous data taking over many months. Recording the temporal behavior of the gas is similarly tedious \cite{Makotyn2014,Klauss2017}.

In this paper we propose and experimentally investigate a scheme for direct observation of second order correlations at short distances within several nanoseconds. 
Since only a very small sample of the gas is observed, most of the gas remains untouched during the measurement and major losses are prevented.
The scheme is based on photoassociative ionization (PI), a technique well known in molecular spectroscopy \cite{Lett1995,Jones2006}.

In PI, a pair of atoms is optically excited to a deeply bound state of a molecular ion. 
The optical excitation is only possible when the distance between the atoms is similar to the size of the bound molecular state. 
The observed ion rate is thus directly proportional to the number of pairs with small interatomic distances.
The ion can be detected with high efficiency.

\begin{figure}[ptb]
\centering\includegraphics[width=1\columnwidth]{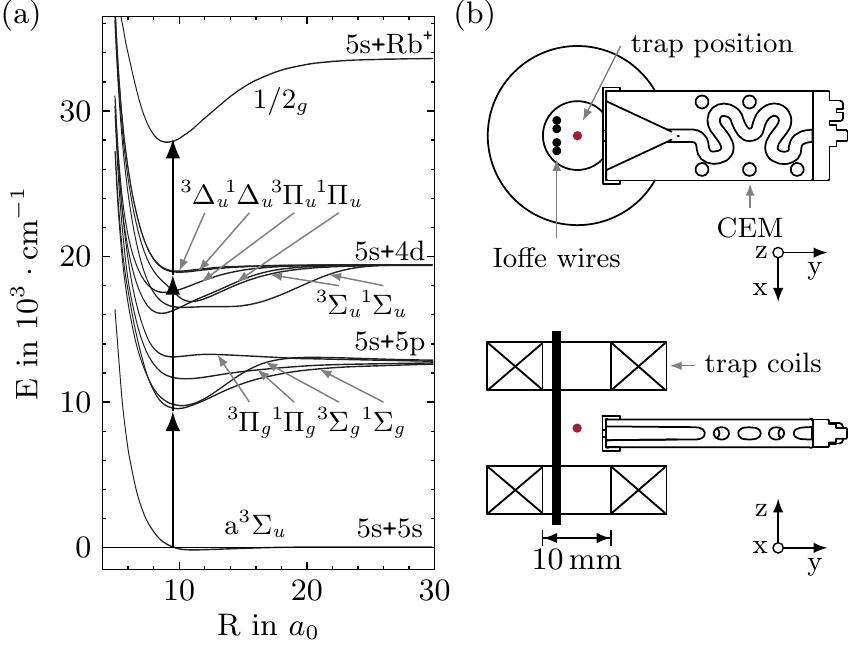}%
\caption{(color online). (a) Molecular potentials relevant for photoassociative
ionization in $^{87}$Rb. 
The internuclear distance $R$ is given in units of Bohr radii $a_{0}$. 
The three vertical arrows indicate a three photon transition from the triplet state of a free pair $^{3}\Sigma_{u}$ to a deeply bound state of the molecular ion in the $1/2_{g}$-potential.
In our case the three photon transition can be induced by the same laser that forms the dipole trap for the atoms.
(b) True to scale cross sectional views of the in vacuum experimental setup with trap coils and channel electron multiplier. 
The outer and inner diameter of the coils are \SI{26}{mm} and \SI{10}{mm}. 
The distance between the coils edges in z-direction is \SI{11}{mm}.}%
\label{fig:1}%
\end{figure}

For $^{87}$Rb atoms the relevant molecular potentials \cite{Jyothi2016,McCabe2010} are shown in Figure \ref{fig:1}(a).
If two atoms in the triplet ground state $^{3}\Sigma_{u}$ approach to a distance of about 10 Bohr radii, a near resonant three photon transition
\cite{Harter2013,*Wolf2017} excites them to a rovibrational state of the $1/2_{g}$-potential of the molecular ion. 
The transition is enhanced by near resonant rovibrational states of various potentials of the 5s+5p asymptote ($^{3}\Pi_{g}$, $^{1}\Pi_{g}$, $^{3}\Sigma_{g}$, $^{1}\Sigma_{g}$) and the 5s+4d asymptote ($^{3}\Delta_{u},^{1}\Delta_{u},^{3}\Pi_{u},^{1}\Pi_{u},^{3}\Sigma_{u},^{1}\Sigma_{u}$). 
The so generated molecular ion is accelerated by an electric field and detected with a channel electron multiplier (CEM) \cite{Gunther2009,Henkel2010}.
We test this scheme with well known Feshbach resonances of cold $^{87}$Rb \cite{Marte2002}.
At resonance, the probability of finding two atoms at a small distance is greatest and the ion counting rate has a maximum \cite{Chin2010,Courteille1998} (see also supplemental information \footnote{See Supplemental Material [url] for the connection between the second order correlation function and the pair wave function, which includes Refs. \cite{Naraschewski1999,Cherny2000,Naidon2003}.}).

The experimental setup that we use is described in more detail in \cite{Marzok2009, Maier2015}. 
In brief, $^{87}$Rb atoms from a magneto-optical trap are magnetically transferred into a Ioffe-type magnetic trap and further cooled by microwave-induced forced evaporation. 
The atoms are then loaded into a crossed optical dipole trap placed at the center of the coils of the Ioffe trap. 
For the dipole trap, we use a linearly polarized multi mode fiber laser with a wavelength near \SI{1070}{\nano\metre} and a bandwidth of \SI{1.8}{nm}. 
The laser output is split into two beams which intersect horizontally under an angle of \SI{36}{\degree} with perpendicular polarizations. 
The beam waist ($1/e^2$-radius) and the total optical power amounts to $w_{0}=\SI{150}{\micro\meter}$ and $P=$ $2\cdot\SI{7}{\watt}$ resulting in trapping frequencies of $\nu_x = \SI{50}{Hz}$, $\nu_y = \SI{153}{Hz}$, $\nu_z = \SI{159}{Hz}$. 
With $5\cdot10^{5}$ atoms at a temperature of \SI{2.8}{\micro\kelvin} the mean number density is \SI{6.7e11}{\centi\metre^{-3}}. 
The optical field of the dipole laser in the trap also drives the above described three photon photoionization. 
After inverting the current in one of the trap coils the atoms are exposed to a homogeneous magnetic field $B$. 
With a current of \SI{35}{A} in both coils one obtains a maximum magnetic field of \SI{1200}{G}. 
The field strength at the position of the atoms was calibrated by microwave spectroscopy of the hyperfine transitions for several current values. 
After preparing the atoms in the hyperfine state $|F,m_{F}\rangle=|1,1\rangle$ by rapid adiabatic passage, the magnetic field is finally ramped up close to a Feshbach resonance. 
The position of the ion detector is shown in figure \ref{fig:1}(b). 
With outer dimensions of $4\times13\times\SI{30}{mm}$, the CEM \footnote{KBL 210, Dr. Sjuts Optotechnik GmbH.} itself is very compact and can be placed between the two trap coils with a distance to the atoms of only \SI{3.7}{mm}. 
Its front aperture is set to the optimum gain voltage of \SI{-2450}{V} and the ions are extracted from the atomic cloud with an electric field of \SI{3.0}{kV/cm}. 
This geometry avoids complicated ion optics and allows to extract ions in magnetic fields of up to \SI{1200}{G} with an detection efficiency of $\varepsilon = 70\%$.

\begin{figure}[ptb]
\centering\includegraphics[width=1\columnwidth]{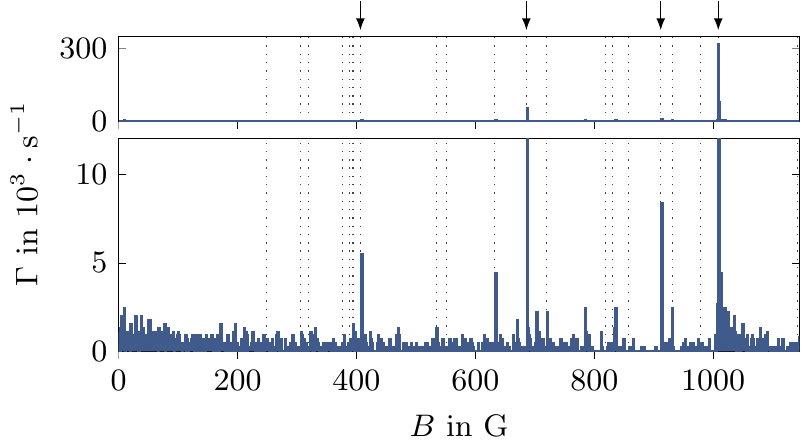}%
\caption{(color online). Observed count rate $\Gamma$ while the magnetic field $B$ is varied from 0 to \SI{1145}{G} within \SI{50}{ms}. 
The plot shows the average over fifty experimental cycles. 
Counts are binned in \SI{44}{\micro\second} intervals. 
Theoretical positions of magnetic s-wave and d-wave Feshbach resonances with entrance channel $|1,1\rangle\times|1,1\rangle$ are indicated with dotted lines \cite{Marte2002}.
The four s-wave resonances are additionally marked with vertical arrows.}%
\label{fig:2}%
\end{figure}

In a first experimental test, the magnetic field is ramped from \SI{0}{G} to \SI{1145}{G} within \SI{50}{ms} and the count rate $\Gamma$ of the ion detector is recorded. 
Figure \ref{fig:2} shows the average over fifty such scans, with each scan carried out in a new experimental cycle. 
One observes a number of sharp peaks with count rates of more then $10^{5}\,\textrm{s}^{-1}$. 
Almost all the peaks are in good agreement with known positions of magnetic s-wave and d-wave Feshbach resonances with entrance channel $|1,1\rangle\times|1,1\rangle$ \cite{Marte2002}. 
For instance, the four most prominent s-wave resonances at \SI{408.9}{G}, \SI{687.1}{G}, \SI{913.3}{G}, \SI{1008.4}{G} match with the theoretical predictions at \SI{406.6}{G}, \SI{685.8}{G}, \SI{911.7}{G}, \SI{1008.5}{G}. 
The lower part of figure \ref{fig:2} with expanded y-axis reveals more resonances which are not yet identified in literature for this entrance channel as for instance at \SI{785.2}{G} and \SI{703.1}{G}. 
While tuning across even the strongest resonance, less than 15 ions are generated such that a single resonance can be detected with almost no atomic losses. 
Apparently, a change of the atomic correlation can be observed in real time within one experimental cycle without significant atom losses. 

\begin{figure}[ptb]
\centering\includegraphics[width=1\columnwidth]{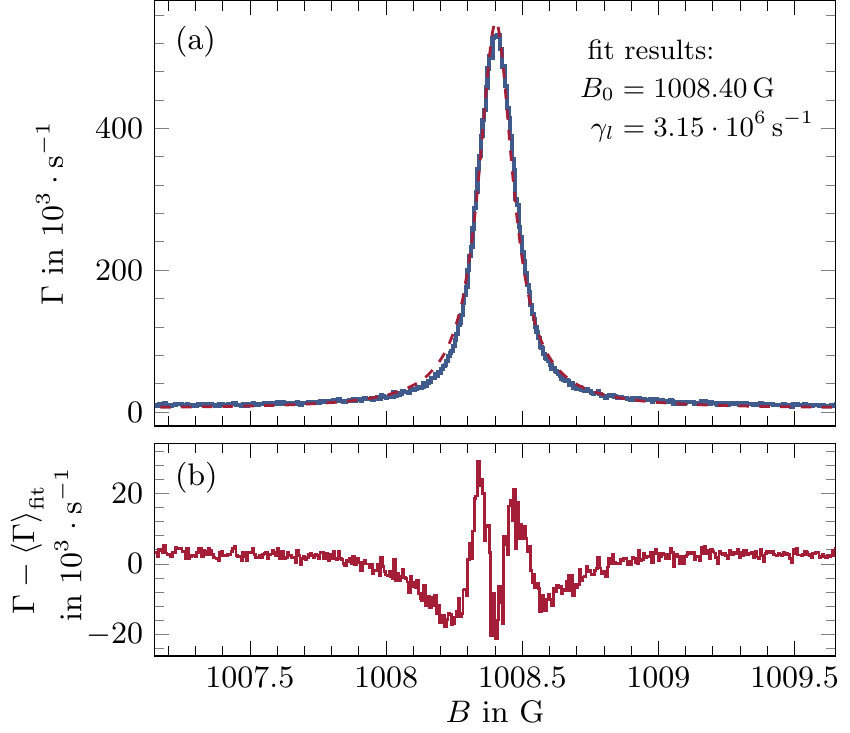}%
\caption{(color online). (a) Feshbach resonance at \SI{1008.4}{G}. 
The plot shows the average over 50 combined up-down-ramps and 100 experimental cycles, with a time binning of \SI{2}{\micro\second}.
The red dashed line shows a theoretical fit according to equation \ref{eq:1}. 
(b) Residual of the fit.}%
\label{fig:3}%
\end{figure}

Due to the small losses it is possible to average over many scans within one experimental cycle. 
In the following we concentrate on the resonance at \SI{1008.5}{G} with highest measured ion rate and largest theoretical width $\Delta=\SI{170}{mG}$ \cite{Marte2002}. 
Within \SI{20}{\milli\second}, we ramp the magnetic field to \SI{1006.7}{G} slightly below the resonance position and wait \SI{15}{\milli\second} for possible perturbations to decay. 
After this relaxation time, we scan the magnetic field 100 times across the resonance with a scan range of \SI{3}{G}. 
Up- and down-ramps are symmetric and each combined up-down-ramp lasts \SI{2}{\milli\second}. 
We repeat this procedure for 100 experimental cycles.
Figure \ref{fig:3}(a) shows the average over 50 combined up-down-ramps and 100 experimental cycles. 
The total data have been recorded within \SI{90}{\minute}. 
The signal-to-noise ratio amounts to about 130. 

The single resonances within this average not all appear at the same magnetic field. 
They are slightly shifted relative to each other due to systematic effects. 
Possible candidates are thermalization of the trap coils and oscillating stray magnetic fields due to power line hum. 
Since complete resonances are recorded with a time separation of only 1ms such slow perturbations can be directly observed by analyzing the position of the resonances. 
The plot in figure \ref{fig:3}(a) is corrected for such systematic shifts. 
The correction algorithm models the perturbations with a test function whose parameters are determined by minimizing the variance of the relative shift of the resonance. 
For details we refer to the supplemental material. 

It is possible to extract the optical excitation rate by comparing the observed resonance line shape with the theoretical expectation. 
The observed ion rate $\Gamma$ may be written as the product of the atom number $N$, the inelastic scattering cross section $\sigma_{i}$, the atom flux $j$ and the CEM detection efficiency $\varepsilon$,
\[
\Gamma=\eta\cdot\varepsilon\cdot N \cdot\sigma_{i}\cdot j.
\]
The factor $\eta$ takes care of the possibility that losses are not exclusively caused by photoassociative ionization (PI) but also by photoassociation (PA). 
Single- and two-photon excitations into rovibrational levels of the neutral molecule (5s+5p and 5s+4d asymptotes) cannot be neglected. 
We thus define the ionization efficiency $\eta$ as the ratio of the rate for PI and the total rate of PI and PA together.
For Feshbach resonances the magnetic field dependence of $\sigma_{i}$ is well known (see for instance section II in ref \cite{Chin2010}) and after averaging over the thermal velocity distribution, we arrive at 
\begin{equation}
\left\langle \Gamma\right\rangle = \eta \cdot \varepsilon \cdot N \cdot g_{\alpha}\cdot\frac{4\pi}{k_{th}^{2}}\cdot n\cdot v_{th}\cdot R\left(B\right).
\label{eq:1}
\end{equation}
In our case of indistinguishable bosons the quantum statistical factor $g_{\alpha}=2$. 
The thermal wave number $k_{th}=$ $\sqrt{2}\mu v_{th}/\hbar$ depends on the thermal velocity $v_{th}=\sqrt{k_{B}T/\mu}$ and the reduced mass $\mu=M_{Rb}/2$. 
The mass of the rubidium atom, the temperature of the gas, the mean number density, and the reduced Planck constant are denoted $M_{Rb}$, $T$, $n$, and $\hbar$. 
The scattering resonance and the thermal averaging is absorbed in the function
\[
R(B)=I\cdot S\int_{0}^{\infty}e^{-\tilde{v}^{2}}\tilde{v}^{2}\frac{1}{(\tilde{v}S+I)^{2}+(B-B_{0})^{2}}d\tilde{v},
\]
with normalized velocity $\tilde{v}=v/v_{th}$, the parameter $I=\hbar \gamma_{l}/(2\mu_{m})$ that describes the ionization and photoassociation losses, the scaled width parameter $S=\mu v_{th}a_{bg}\Delta/\hbar$, and the resonance position $B_{0}$. 
The differential magnetic moment $\mu_{m}$, the background scattering length $a_{bg}$ and the resonance width $\Delta$ are standard parameters of Feshbach resonances and are listed in \cite{Chin2010}. 
For our resonance $\mu_{m}=2.79\cdot\mu_{B}$, $a_{bg}=100\cdot a_{0}$, and $\Delta=\SI{170}{mG}$ with Bohr magneton $\mu_{B}$ and Bohr radius $a_{0}$. 
The loss parameter $\gamma_{l}$ describes the rate at which the population of the stationary scattering state decays due to losses. 
In our case the total loss rate is the sum $\gamma_{l}=\gamma_{pi}+\gamma_{pa}$ of the optical three-photon excitation rate $\gamma_{pi} $ and the optical photoassociation rate $\gamma_{pa}$.
Because of the large bandwidth of the excitation laser dopplershifts can be neglected and $\gamma_{l}$ does not depend on velocity.
The ionization efficiency can be expressed in terms of the loss parameters as $\eta =\gamma _{pi}/\gamma _{l}$.

A fit to the data shown in figure \ref{fig:3}(a) yields $B_{0}=\SI{1008.4}{G}$ and $\gamma_{l}=\SI{3.15e6}{s^{-1}}$. In figure \ref{fig:3}(b) the difference between the fit and the measured data is shown. 
The systematic discrepancy of about 4\% might be due to magnetic field noise that slightly smears out the resonance and changes its observed line shape. 
Furthermore, in the model the light intensity and the number density is assumed to be homogeneous. 
Fully understanding the residual requires a numerical simulation of the gas in a trapping potential with finite depth including a noisy magnetic field and a laser field with exactly known intensity profile. 
In view of its complexity we refrain from such a program at this point.

\begin{figure}[ptb]
\centering\includegraphics[width=1\columnwidth]{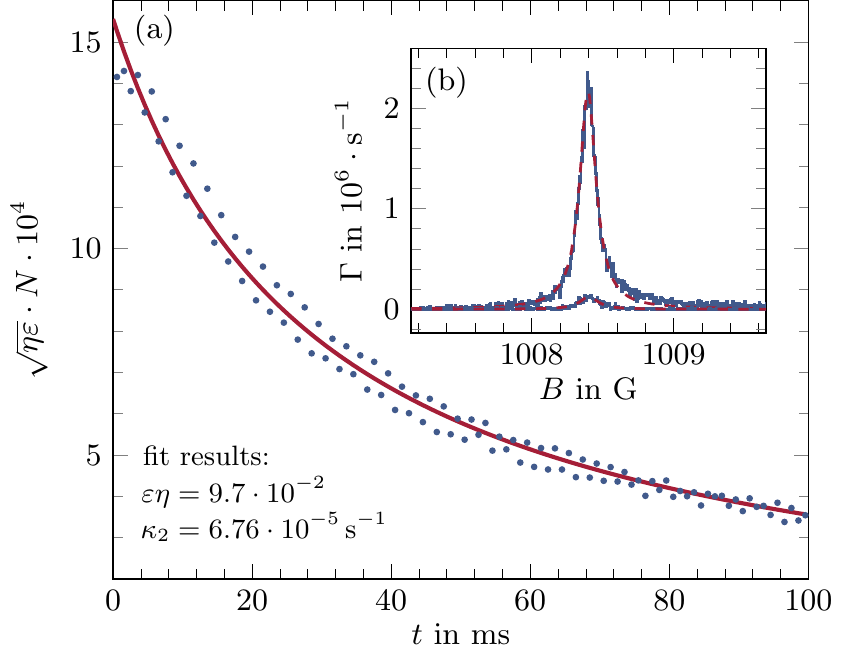}%
\caption{(color online). (a) Decay of the atom number due to optical losses. 
The red line shows the theoretical expectation for $\varepsilon\eta=\SI{9.7e-2}{}$. 
For details see text. 
The atom number differs systematically depending on the scan direction of the magnetic field. 
This effect leads to a deviation in the fit results of below ten percent.
Its origin is yet unclear and requires further investigation.
(b) Resonances corresponding to the first and last point of the decay curve, demonstrating the high data quality even at the end of the decay interval. 
The red lines are fits of the theoretical model to the measured data analogous to figure \ref{fig:3}(a). }%
\label{fig:4}%
\end{figure}

We can extract further information, if we sort the data such that the decay of the gas becomes visible. 
To this end, we average each ramp individually over the hundred experimental cycles. 
One obtains 100 resonances taken in time steps of 1ms. 
A fit of each resonance according to equation \ref{eq:1} with $\eta\cdot\varepsilon\cdot N^{2}$ as fit parameter leads to the decay curve $\sqrt{\eta\varepsilon} \cdot N(t)$ shown in figure \ref{fig:4}(a) \footnote{Note that $N\cdot n=N^{2}\cdot n/N$ with $n/N$ being constant.}.

We observe a decay of the gas within a few ten milliseconds.
This decay is too fast to be explained by three-body collisions which for our density would last several seconds \cite{Marte2002}. 
If this decay is caused by PI and PA, the temporal change of the atom number is given by twice the total optical excitation rate, $\dot{N}=-2\ \cdot\left\langle \Gamma\right\rangle /(\varepsilon\eta)$. 
The factor $2$ takes into account that the count rate is proportional to atom pairs. 
With $\Gamma$ from equation \ref{eq:1} one obtains $\dot{N}=-\kappa_{2}N^{2}$ with decay coefficient $\kappa_{2}=2\cdot g_{\alpha}\cdot4\pi/k_{th}^{2}\cdot n/N\cdot v_{th}\cdot\overline{R}$. 
At each time step the magnetic field is tuned across the resonance such that only the average over the scan range $\overline{R}=\int_{B_{\text{min}}}^{B_{\text{max}}}{R(B)dB}/(B_{\text{max}}-B_{\text{min}})$ is relevant for the losses. 
The equation for $N$ is solved by $N(t)=N_{0}/(t\kappa_{2} N_{0}+1)$. 
By fitting the theoretical expectation $\sqrt{\eta\varepsilon}\cdot N\left(t\right)$ to the data shown in figure \ref{fig:4}(a), we obtain $\eta\varepsilon=0.097$. 
The initial atom number $N_{0}=5\cdot10^{5}$ is determined by conventional absorption imaging. 
With $\varepsilon=0.7$ one obtains $\eta=0.14$ and $\gamma _{pi}=\eta\cdot\gamma_{l}=0.14\cdot\SI{3.15e6}{s^{-1}}=\SI{4.4e5}{s^{-1}}$. 
Obviously, the photoassociation rate $\gamma_{pa}=(1-\eta)\gamma_{l}=\SI{2.71e6}{s^{-1}}$ dominates the losses in our specific scenario. 
The fit also yields a value for $\kappa_{2}=\SI{6.76e-5}{s^{-1}}$ which is greater than the theoretically expected value of \SI{1.58e-5}{s^{-1}} but within the same order of magnitude.
The discrepancy may be due to density fluctuations \cite{Burt1997} and to an overestimation of the effective trapping volume which cannot be determined very precisely from the observed trap oscillation frequencies. 

Although the atom losses are small, each optical excitation destroys a strongly correlated atom pair. 
In the case of a Feshbach resonance this effect is included in the above model \cite{Chin2010}.
For large $I$ the complex scattering phase
\[
\varphi=-\arctan\left(ka_{bg}\right)-\arctan\left(ka_{bg}\Delta\frac{B-B_{0}+i\cdot I}{\left(B-B_{0}\right)^{2}+I^{2}}\right)
\]
approaches $-\arctan\left(ka_{bg}\right)$ and the scattering resonance disappears. 
The correlation properties of the gas remain unchanged by the detection, only if the rate at which correlated pairs are generated exceeds the optical loss rate. 
It is thus desirable to minimize any unwanted photoassociation. 
However, controlled optical losses may also be used to modify second order correlations. 
In our experiment the losses are so strong that the maximum elastic scattering length $a=-\operatorname{Re}(\tan\varphi)/k$ exceeds the background scattering length $a_{bg}$ by only a factor of about two. 
Obviously, it is possible to detect and ramp across a Feshbach resonance without significantly changing the elastic scattering properties and the related atomic interaction in the gas. 

In summary we have shown that photoassociative ionization provides a powerful tool for studying atomic Feschbach resonances and the related change of the pair correlation. 
Changes in the correlation can be monitored within several ten nanoseconds limited only by the optical excitation time and the raise time of the CEM. 
This opens up new possibilities to investigate the quench behavior of quantum gases in various scenarios \cite{Fletcher2017} but it also may allow to push the limits of few-body physics. 
Real time detection of heteronuclear Efimov trimers \cite{Maier2015,Ulmanis2016,Johansen2017} might be possible for instance by looking at the correlation of the two heavy atoms. 
Near a heteronuclear Feshbach resonance a third light atom binds the two heavy atoms with an additional $1/R^{2}$ potential. 
The enhanced correlation of the heavy atoms at small distances would show up immediately in the ion count rate. 
The method may also help to find new and weak Feshbach resonances in survey experiments \cite{Sawyer2017} that exploit the new possibilities of wide and fast scans. 
Even in experiments where the correlation peaks at larger distances as for instance in optical lattices, the method might be useful since the enhanced correlation at large distances is related to a reduced correlation at small distances and the ion signal should drop.

Concerning technical improvement, it would be desirable to separate the optical excitation from the dipole trap. 
If the dipole trap is formed with light that does not ionize the gas, an extra laser may be introduced that is optimized for ionization with two- or even single-photon transitions. 
This would suppress unwanted photoassociation losses since the intermediate molecular levels would be far detuned. 
Furthermore, it would be interesting to find PI schemes for other species and extend the method even to non alkali quantum gases. 
From a theoretical point of view it may be interesting to better understand the connection between the ionization rate and the states of the atomic ensemble in Mott insulator transitions, BEC-BCS cross over, Bose-Einstein condensation, supersolidity in dipolar gases, low dimensional systems, or systems with cavity mediated long range interaction. 
Finally, our experiment shows that it is possible to efficiently generate cold molecular ions by photoassociative ionization in combination with Feshbach resonances \cite{Jyothi2016}.

\begin{acknowledgments}
We acknowledge helpful discussions with Andreas G\"{u}nther and Peter Federsel. The work has been supported by the Deutsche Forschungsgemeinschaft (Zi 419/8-1).
\end{acknowledgments}


\providecommand{\noopsort}[1]{}\providecommand{\singleletter}[1]{#1}%

\end{document}